# The PBD-Closure of Constant-Composition Codes

Yeow Meng Chee, Alan C. H. Ling, San Ling, and Hao Shen

*Abstract*—We show an interesting pairwise balanced design (PBD)-closure result for the set of lengths of constant-composition codes whose distance and size meet certain conditions. A consequence of this PBD-closure result is that the size of optimal constant-composition codes can be determined for infinite families of parameter sets from just a single example of an optimal code. As an application, the size of several infinite families of optimal constant-composition codes are derived. In particular, the problem of determining the size of optimal constant-composition codes having distance four and weight three is solved for all lengths sufficiently large. This problem was previously unresolved for odd lengths, except for lengths seven and eleven.

## I. INTRODUCTION

ONE generalization of constant-weight binary codes as we enlarge the alphabet from $\mathbb{Z}_2$ to $\mathbb{Z}_q$ ($q > 2$) is the concept of constant-composition codes. The class of constant-composition codes includes the important permutation codes and has attracted recent interest due to its numerous applications, such as in determining the zero error decision feedback capacity of discrete memoryless channels [1], multiple-access communications [2], spherical codes for modulation [3], DNA codes [4], [5], powerline communications [6], [7], and frequency hopping [8].

While constant-composition codes have been used since the early 1980s to bound error and erasure probabilities in decision feedback channels [9], their systematic study only began in late 1990s with Svanström [10]. Today, the problem of determining the maximum size of a constant-composition code constitutes a central problem in their investigation [6], [7], [11]–[19].

A $q$-ary code of length $n$ is a set $\mathcal{C} \subseteq \mathbb{Z}_q^n$. The *Hamming distance* between two codewords $\mathsf{u}, \mathsf{v} \in \mathcal{C}$ is denoted by $d_H(\mathsf{u}, \mathsf{v})$. A code $\mathcal{C}$ is said to have *distance* $d$ if $d_H(\mathsf{u}, \mathsf{v}) \geq d$ for all $\mathsf{u}, \mathsf{v} \in \mathcal{C}$. The *weight* of a codeword $\mathsf{u} \in \mathcal{C}$ is the number of nonzero components of $\mathsf{u}$. If every codeword in $\mathcal{C}$ has weight $w$,

Manuscript received October 24, 2006; revised April 7, 2007. The research of Y. M. Chee and S. Ling was supported in part by the Singapore Ministry of Education under Research Grant T206B2204.

Y. M. Chee is with the Interactive Digital Media R&D Program Office, Media Development Authority, Singapore 179369. He is also with the Division of Mathematical Sciences, School of Physical and Mathematical Sciences, Nanyang Technological University, Singapore 637616 (e-mail: ymchee@alumni.uwaterloo.ca).

A. C. H. Ling is with the Department of Computer Science, University of Vermont, Burlington, VT 05405 USA (e-mail: aling@emba.uvm.edu).

S. Ling is with the Division of Mathematical Sciences, School of Physical and Mathematical Sciences, Nanyang Technological University, Singapore 637616 (e-mail: lingsan@ntu.edu.sg).

H. Shen is with the Department of Mathematics, Shanghai Jiao Tong University, Shanghai 200240, China (e-mail: haoshen@sjtu.edu.cn).

Communicated by G. Seroussi, Associate Editor for Coding Theory.

Digital Object Identifier 10.1109/TIT.2007.901175

then $\mathcal{C}$ is said to be of (constant) *weight* $w$. The *composition* of a codeword $\mathsf{u} \in \mathcal{C}$ is the tuple $\bar{w} = [w_1, \ldots, w_{q-1}]$ such that $\mathsf{u}$ contains exactly $w_i$ occurrences of $i, i \in \mathbb{Z}_q \setminus \{0\}$. A $q$-ary code $\mathcal{C}$ has (constant) *composition* $\bar{w}$ if every codeword in $\mathcal{C}$ has composition $\bar{w}$. A $q$-ary code of length $n$, distance $d$, and composition $\bar{w}$ is referred to as an $(n, d, \bar{w})_q$-code. The maximum size of an $(n, d, \bar{w})_q$-code is denoted $A_q(n, d, \bar{w})$ and the $(n, d, \bar{w})_q$-codes achieving this size are called *optimal*. Note that the following operations do not affect distance and weight properties of an $(n, d, \bar{w})_q$-code:

 i) reordering the components of $\bar{w}$, and
 ii) deleting zero components of $\bar{w}$.

Consequently, throughout this paper, we restrict our attention to those $\bar{w} = [w_1, \ldots, w_{q-1}]$, where $w_1 \geq \cdots \geq w_{q-1} \geq 1$.

We note that our notation for constant-composition codes differs from that in the literature in that we do not explicitly specify $w_0$, the number of occurrences of zero in each codeword. The reason is that $w_0$ can be inferred from $n$ and $w_1, \ldots, w_{q-1}$, so there is no need to specify $w_0$. Besides the advantage of being more succinct, our notation for constant-composition codes seems more natural and convenient for the investigation of their combinatorial properties.

Our starting point is the problem of determining the size of optimal ternary constant-composition codes of distance four and weight three, that is, the determination of $A_3(n, 4, [2, 1])$. This problem was first investigated by Svanström [11], who determined $A_3(n, 4, [2, 1])$ for all even $n$ as well as for $n = 7$ and $n = 11$. The value of $A_3(n, 4, [2, 1])$ remains unknown for all other odd $n$. In this paper, we develop a technique for determining $A_3(n, 4, [2, 1])$ for all but a finite set of values of odd $n$. We do this by showing that the set of lengths of constant-composition codes satisfying certain size and distance constraints is pairwise balanced design (PBD) closed.

It turns out this PBD-closure property is a general result applicable to the determination of $A_q(n, d, \bar{w})$ beyond the case of ternary constant-composition codes of weight three. It allows us to bring to bear Wilson's powerful theory of PBD-closure [20]–[22], which has been a cornerstone in the development of modern combinatorial design theory. One consequence is that for any fixed $q, d$, and $\bar{w}$ satisfying certain conditions, $A_q(n, d, \bar{w})$ can be determined for infinitely many $n$ from just a single example of an optimal $(n_0, d, \bar{w})_q$-code. We illustrate the applicability of our PBD-closure result by also determining the size of several infinite families of optimal constant-composition codes.

## II. PBD-CLOSURE AND RECURSIVE CONSTRUCTIONS

For $n$ a positive integer, the set $\{1, 2, \ldots, n\}$ is denoted $[n]$. The $i$th component of a vector $\mathsf{u}$ is denoted $\mathsf{u}_i$. All vectors considered in this paper have components indexed from zero, meaning $(u_0, u_1, \ldots, u_{n-1})$. The *support* of a vector $\mathsf{u}$ is the set $\mathrm{supp}(\mathsf{u}) = \{i : \mathsf{u}_i \neq 0\}$.

A *set system* is a pair $\mathcal{S} = (X, \mathcal{A})$, where $X$ is a finite set and $\mathcal{A} \subseteq 2^X$. The members of $X$ are called *points* and the members of $\mathcal{A}$ are called *blocks*. The *order* of $\mathcal{S}$ is the number of points $|X|$, and the *size* of $\mathcal{S}$ is the number of blocks $|\mathcal{A}|$. A set $K$ is called a *set of block sizes* for $\mathcal{S}$ if $|A| \in K$ for all $A \in \mathcal{A}$. $\mathcal{S}$ is *k-uniform* if $\{k\}$ is a set of block sizes for $\mathcal{S}$.

A *pairwise balanced design* (PBD) is a set system $(X, \mathcal{A})$ such that every 2-subset of $X$ is contained in exactly one block. A PBD of order $n$ with set of block sizes $K$ is denoted $\mathrm{PBD}(n, K)$. A set $S$ of positive integers is *PBD-closed* if the existence of a $\mathrm{PBD}(n, S)$ implies that $n \in S$. Given a set $K$ of positive integers, the *PBD-closure* of $K$ is the set

$$B(K) = \{n : \text{there exists a } \mathrm{PBD}(n, K)\}.$$

Let $c$ be a positive real number and define

$$M_c(d, \bar{w}; q) = \{n : A_q(n, d, \bar{w}) \geq cn(n-1)\}.$$

*Theorem 1:* Let $\bar{w} = [w_1, \ldots, w_{q-1}]$ and $w = \sum_{i=1}^{q-1} w_i$. The set $M_c(d, \bar{w}; q)$ is PBD-closed, provided $d \leq 2w - 2$.

*Proof:* Suppose $(\mathbb{Z}_n, \mathcal{A})$ is a $\mathrm{PBD}(n, M_c(d, \bar{w}; q))$. We construct an $(n, d, \bar{w})_q$-code as follows. For each block $A \in \mathcal{A}$, we put an $(n, d, \bar{w})_q$-code of size at least $c|A|(|A|-1)$ on $A$: the codewords are length $n$ vectors so that every position $i \notin A$ has value zero, and when restricted to the positions $i \in A$, the codewords form an $(|A|, d, \bar{w})_q$-code of size at least $c|A|(|A|-1)$. An $(|A|, d, \bar{w})_q$-code of size at least $c|A|(|A|-1)$ exists since $|A| \in M_c(d, \bar{w}; q)$. We claim that the resulting code is an $(n, d, \bar{w})_q$-code. Indeed, any two codewords arising from the same block in the PBD are distance at least $d$ apart. So we need only check the distance between codewords arising from different blocks. Since any two blocks intersect in at most one point, the supports of two codewords arising from two different blocks intersect in at most one point. So these two codewords must be at least distance $2w - 2$ apart. Since $d \leq 2w - 2$, the resulting code is an $(n, d, \bar{w})_q$-code. It remains to compute the size of this code.

Let $b_k$ denote the number of blocks of size $k$ in $\mathcal{A}$. Since every 2-subset of $\mathbb{Z}_n$ is contained in exactly one block, we have

$$\binom{n}{2} = \sum_{k \in M_c(d, \bar{w}; q)} b_k \binom{k}{2}. \tag{1}$$

Now, on each block of size $k$, we placed a code of size at least $ck(k-1)$. So the size of the resulting code we constructed is at least

$$\sum_{k \in M_c(d, \bar{w}; q)} b_k ck(k-1).$$

By comparing with (1), we see that this quantity is equal to $cn(n-1)$. Consequently, $n \in M_c(d, \bar{w}; q)$. Hence, $M_c(d, \bar{w}; q)$ is PBD-closed. □

For a positive constant $c$, and $n \notin M_c(n, d, \bar{w}; q)$, the construction in the proof of Theorem 1 still yields a constant-composition code, although its size cannot be determined without more precise knowledge of the number of blocks of each size in the PBD. To have this additional knowledge, we focus on PBDs coming from the class of *group divisible designs*.

Let $(X, \mathcal{A})$ be a set system and let $\mathcal{G} = \{G_1, \ldots, G_s\}$ be a partition of $X$ into subsets, called *groups*. The triple $(X, \mathcal{G}, \mathcal{A})$ is a *group divisible design* (GDD) when every 2-subset of $X$ not contained in a group appears in exactly one block, and $|A \cap G| \leq 1$ for all $A \in \mathcal{A}$ and $G \in \mathcal{G}$. We denote a GDD $(X, \mathcal{G}, \mathcal{A})$ by $K$-GDD if $K$ is a set of block sizes for $(X, \mathcal{A})$. The *type* of a GDD $(X, \mathcal{G}, \mathcal{A})$ is the multiset $\{|G| : G \in \mathcal{G}\}$. When more convenient, we use the exponentiation notation to describe the type of a GDD: a GDD of type $g_1^{t_1} \cdots g_s^{t_s}$ is a GDD where there are exactly $t_i$ groups of size $g_i, i \in [s]$.

PBDs and GDDs are intimately related. A $\mathrm{PBD}(n, K)$ is a $K$-GDD of type $1^n$. Given a $K$-GDD $(X, \mathcal{G}, \mathcal{A})$ of type $g_1^{t_1} \cdots g_s^{t_s}$, we can treat it as a $\mathrm{PBD}(\sum_{i \in [s]} g_i t_i, K')$, where $K' = K \cup \{g_i : i \in [s]\}$, by considering the groups also as blocks.

We now extract the construction in the proof of Theorem 1 and apply it to GDDs.

*Theorem 2:* Suppose there is a $K$-GDD of type $g_1^{t_1} \cdots g_s^{t_s}$ containing $b_k$ blocks of size $k \in K$. Let $n = \sum_{i=1}^{s} g_i t_i$, $\bar{w} = [w_1, \ldots, w_{q-1}]$, and $w = \sum_{i=1}^{q-1} w_i$. Then

$$A_q(n, d, \bar{w}) \geq \sum_{k \in K} b_k A_q(k, d, \bar{w}) + \sum_{i \in [s]} t_i A_q(g_i, d, \bar{w})$$

provided $d \leq 2w - 2$.

### III. USING PBD-CLOSURE TO DETERMINE $A_q(n, d, \bar{w})$

The PBD-closure result of Theorem 1 is particularly useful in determining $A_q(n, d, \bar{w})$ of the form $cn(n-1)$, for some positive constant $c$.

*Theorem 3:* Let $\bar{w} = [w_1, \ldots, w_{q-1}], w = \sum_{i=1}^{q-1} w_i$ and $d \leq 2w - 2$. Suppose for some constant $c$, $A_q(n, d, \bar{w}) \leq cn(n-1)$. Then whenever we can find a set $K$ such that there exists a $(k, d, \bar{w})_q$-code of size $ck(k-1)$ for each $k \in K$, we have $A_q(n, d, \bar{w}) = cn(n-1)$ for all $n \in B(K)$.

*Proof:* By Theorem 1, there exists an $(n, d, \bar{w})_q$-code of size at least $cn(n-1)$ for each $n \in B(K)$. This combined with the upper bound $A_q(n, d, \bar{w}) \leq cn(n-1)$ implies that $A_q(n, d, \bar{w}) = cn(n-1)$. □

The usefulness of Theorem 3 lies in Wilson's theory of asymptotic existence of PBDs, which essentially says that the obvious necessary conditions for the existence of a PBD are asymptotically sufficient. More formally, we have the following.

*Theorem 4 (Wilson [22]):* Let $K$ be a set of positive integers. Define

$$\alpha(K) = \gcd\{k - 1 : k \in K\}$$
$$\beta(K) = \gcd\{k(k-1) : k \in K\}.$$

Then there is a constant $n_0(K)$ such that for all $n \geq n_0(K)$ satisfying

$$n - 1 \equiv 0 \pmod{\alpha(K)}$$

and

$$n(n-1) \equiv 0 \pmod{\beta(K)}$$

we have $n \in B(K)$.

TABLE I
NECESSARY CONDITIONS FOR THE EXISTENCE OF $\{5\}$-GDD OF TYPE $g^u$

| $g \pmod{20}$ | Condition on $u$ |
|---|---|
| 0 | $u \geq 5$ |
| 1, 3, 7, 9, 11, 13, 17, 19 | $u \equiv 1$ or $5 \pmod{20}$ |
| 2, 6, 14, 18 | $u \equiv 1$ or $5 \pmod{10}$ |
| 4, 8, 12, 16 | $u \equiv 0$ or $1 \pmod 5$ |
| 5, 15 | $u \equiv 1 \pmod 4$ |
| 10 | $u \equiv 1 \pmod 2$ and $u \geq 5$ |

Hence, from a small finite number of examples of optimal codes, we can derive infinite families of optimal codes. The following Johnson-type bound has been proven for constant-composition codes.

*Lemma 1 (Svanström et al. [12]):*

$$A_q(n, d, [w_1, \ldots, w_{q-1}]) \leq \frac{n}{w_1} A_q(n-1, d, [w_1-1, \ldots, w_{q-1}]).$$

When $d = 2\sum_{i=1}^{q-1} w_i - 2$, we have

$$A_q(n-1, d, [w_1-1, \ldots, w_{q-1}]) \leq \left\lfloor \frac{n-1}{\sum_{i=1}^{q-1} w_i - 1} \right\rfloor$$

so that the inequality in Lemma 1 takes the form

$$A_q(n, d, [w_1, \ldots, w_{q-1}]) \leq \left\lfloor \frac{n}{w_1} \left\lfloor \frac{n-1}{\sum_{i=1}^{q-1} w_i - 1} \right\rfloor \right\rfloor. \quad (2)$$

When $d = 2\sum_{i=1}^{q-1} w_i - 3$ and $w_1 \geq 2$, by applying Lemma 1, we obtain

$$A_q(n-1, d, [w_1-1, \ldots, w_{q-1}]) \leq \frac{n-1}{w_1 - 1} A_q(n-2, d, [w_1-2, \ldots, w_{q-1}]).$$

But $A_q(n-2, d, [w_1-2, \ldots, w_{q-1}]) = 1$, so that the inequality in Lemma 1 takes the form

$$A_q(n, d, [w_1, \ldots, w_{q-1}]) \leq \left\lfloor \frac{n}{w_1} \left\lfloor \frac{n-1}{w_1 - 1} \right\rfloor \right\rfloor. \quad (3)$$

Hence, the PBD-closure result of Theorem 1 is potentially applicable to the determination of $A_q(n, d, [w_1, \ldots, w_{q-1}])$ when $d = 2\sum_{i=1}^{q-1} w_i - 3$ and $w_1 \geq 2$, and when $d = 2\sum_{i=1}^{q-1} w_i - 2$, since under these conditions, the upper bound on $A_q(n, d, [w_1, \ldots, w_{q-1}])$ has the form $cn(n-1)$ for some positive constant $c$. We provide details below for the determination of several infinite families of $A_q(n, d, \bar{w})$. The $\{4\}$-GDDs and $\{5\}$-GDDs are used extensively to arrive at the results. So we state their existence below for easy reference.

*Theorem 5 (Ge and Rees [23]):* There exists a $\{4\}$-GDD of type $g^4 m^1$ with $m > 0$ if and only if $g \equiv m \equiv 0 \pmod 3$ and $0 < m \leq 3g/2$.

*Theorem 6 (Ge and Ling [24]):* The necessary conditions in Table I for the existence of a $\{5\}$-GDD of type $g^u$ is also sufficient, except when $g^u \in \{2^5, 2^{11}, 3^5, 6^5\}$, and possibly when
  i) $g^u \in \{3^{45}, 3^{65}\}$;
  ii) $g \equiv 2, 6, 14, 18 \pmod{20}$ and
    a) $g = 2$ and
    $u \in \{15, 35, 71, 75, 95, 111, 115, 195, 215\}$;
   b) $g = 6$ and $u \in \{15, 35, 75, 95\}$;
   c) $g = 18$ and $u \in \{11, 15, 71, 111, 115\}$;
   d) 
   $$g \in \{14, 22, 26, 34, 38, 46, 58, 62\}$$
   and
   $$u \in \{11, 15, 71, 75, 111, 115\};$$
   e) $g \in \{42, 54\}$ or $g = 2a$ with $a \equiv 1, 3, 7,$ or $9 \pmod{10}$ and $33 \leq a \leq 2443$, and $u = 15$;
  iii) $g \equiv 10 \pmod{20}$ and
   a) $g = 10$ and $u \in \{5, 7, 15, 23, 27, 33, 35, 39, 47\}$;
   b) $g = 30$ and $u \in \{9, 15\}$;
   c) $g = 50$ and $u \in \{15, 23, 27\}$;
   d) $g = 90$ and $u = 23$;
   e) $g = 10a, a \equiv 1 \pmod 6, 7 \leq a \leq 319$, and $u \in \{15, 23\}$;
   f) $g = 10b, b \equiv 5 \pmod 6, 11 \leq b \leq 443$, and $u \in \{15, 23\}$;
   g) $g = 10c, c \equiv 1 \pmod 6, 325 \leq c \leq 487$, and $u = 15$;
   h) $g = 10d, d \equiv 5 \pmod 6, 449 \leq c \leq 485$, and $u = 15$.

*Lemma 2:* There exists a $\{9\}$-GDD of type $3^{33}$.

*Proof:* A $\{9\}$-GDD of type $3^{33}$ is first found by Mathon (unpublished), arising from an elliptic semiplane. A construction is given in [25]. □

## IV. BOUNDS

Since Theorem 1 is applicable only for codes whose size meets bounds of a specific form, we list here upper bounds for the size of some codes of interest in this paper. Except for the bound on $A_3(n, 4, [2, 1])$ when $n \equiv 3 \pmod 4$, the bounds presented are all consequences of inequalities (2) and (3).

### A. Bounds on $A_3(n, 4, [2, 1])$

Svanström [11] proved that when $n \equiv 1 \pmod 4$

$$A_3(n, 4, [2, 1]) \leq \frac{n(n-1)}{4}$$

and when $n \equiv 3 \pmod 4$

$$A_3(n, 4, [2, 1]) \leq \frac{(n-1)^2}{4} + \left\lfloor \frac{n-3}{12} \right\rfloor.$$

Define

$$\epsilon(n) = \begin{cases} 0, & \text{if } n \equiv 1 \pmod 4 \\ n, & \text{if } n \equiv 3 \pmod{12} \\ n+2, & \text{if } n \equiv 7 \pmod{12} \\ n+4, & \text{if } n \equiv 11 \pmod{12}. \end{cases}$$

We can unify the above bounds as follows.

*Lemma 3:* For $n$ odd

$$A_3(n, 4, [2, 1]) \leq \frac{n(n-1)}{4} - \frac{\epsilon(n)}{6}.$$

### B. Bounds on Some Ternary Constant-Composition Codes of Weight Four and Five

*Lemma 4:*

$$A_3(n,d,\bar{w}) \leq \begin{cases} \frac{n(n-1)}{6}, & \text{if } (d,\bar{w}) = (5,[3,1]), (6,[2,2]), \\ & \qquad (6,[2,1,1]) \\ \frac{n(n-1)}{9}, & \text{if } (d,\bar{w}) = (6,[3,1]) \\ \frac{n(n-1)}{12}, & \text{if } (d,\bar{w}) = (7,[4,1]). \end{cases}$$

### C. Bounds on $A_4(n,d,[1,1,1])$

*Lemma 5:*

$$A_4(n,d,[1,1,1]) \leq \begin{cases} n(n-1), & \text{if } d=3 \\ n\lfloor \frac{n-1}{2} \rfloor, & \text{if } d=4 \\ n, & \text{if } d=5. \end{cases}$$

### D. Refinements

Let $\bar{w} = [w_1, \ldots, w_a]$ and $\bar{v} = [v_1, \ldots, v_b]$. We say that $\bar{w}$ is a *refinement* of $\bar{v}$ if there is a partition $\{I_1, \ldots, I_b\}$ of $\{1, \ldots, a\}$ such that $\sum_{i \in I_j} w_i = v_j$ for each $j$. The following simple observation of Chu *et al.* [8] is sometimes useful in deriving optimal constant-composition codes.

*Lemma 6 (Chu et al. [18]):* If $\bar{w}$ is a refinement of $\bar{v}$, then $A_q(n,d,\bar{w}) \geq A_{q'}(n,d,\bar{v})$.

## V. SHORT CODES

The ingredients we need in order to apply the PBD-closure result and GDD-based constructions are short optimal codes. We give the existence of some of these in this section. The methods used to construct these codes fall into three categories.

  i) Manual or exhaustive search: The really short codes can be constructed by hand or exhaustive search.
  ii) Clique-finding for cyclic codes: Here, the orbits of codewords of length $n$ and composition $\bar{w}$ are represented by vertices of a graph $G$. We disregard orbits that contain two codewords having distance less than $d$. An edge exists between two vertices if and only if the corresponding orbits do not contain two codewords that are of distance less than $d$ apart. The set of orbits corresponding to a clique in $G$ then gives an $(n,d,\bar{w})_q$-code. A maximum clique finding program (`Cliquer` [26]) is used to determine the largest cyclic $(n,d,\bar{w})_q$-code. We then check to see if this code is optimal.
  iii) Stochastic local search where others fail: When an optimal cyclic code does not exist, or if the graph becomes large, the clique-finding approach becomes inapplicable. In this scenario, we use a stochastic local search method to construct the code.

Our results can be summarized as follows.

*Lemma 7:* $A_3(n,4,[2,1]) = n(n-1)/4 - \epsilon(n)/6$ when $n \in \{5,9,13,15,17,19,23,27,29,31,33\}$.

*Lemma 8:* $A_4(n,3,[1,1,1]) = n(n-1)$ when

$$n \in \{4,7,8,9,10,11,12,14,15,17,18,19,20,\\ 21,23,24,26,27,30,35,38,39,41,42\}.$$

*Lemma 9:* $A_4(n,4,[1,1,1]) = n\lfloor (n-1)/2 \rfloor$ when

$$n \in \{4,10,11,14,16,18,19,22,23,24,\\ 25,26,27,28,29,30,31,32,33,34\}.$$

*Lemma 10:* $A_3(n,5,[3,1]) = n(n-1)/6$ when $n \in \{7,13,15,19\}$.

*Lemma 11:* $A_3(n,7,[4,1]) = n(n-1)/12$ when $n \in \{13,16\}$.

*Lemma 12:* $A_4(5,3,[1,1,1]) = 18$ and $A_4(6,3,[1,1,1]) = 28$.

All the optimal codes proving the lemmas above can be obtained from the first author's website at

⟨http://www1.spms.ntu.edu.sg/~ymchee/ccc.php⟩

The optimality of the codes in Lemmas 7–11 comes from the upper bounds given in Section IV. Optimality of the codes in Lemma 12 is established by exhaustive search (see also [15]).

## VI. OPTIMAL CONSTANT-COMPOSITION CODES OF WEIGHT $W$ AND DISTANCE $2w - 2$

Let $\bar{w} = [w_1, \ldots, w_{q-1}], w = \sum_{i=1}^{q-1} w_i$ and $d = 2w-2$. Lemma 1 gives

$$A_q(n,d,\bar{w}) \leq \left\lfloor \frac{n}{w_1} \left\lfloor \frac{n-1}{w-1} \right\rfloor \right\rfloor.$$

Hence, when $n$ satisfies

$$n - 1 \equiv 0 \pmod{w-1}$$

and

$$n(n-1) \equiv 0 \pmod{w_1(w-1)}$$

we have

$$A_q(n,d,\bar{w}) \leq \frac{n(n-1)}{w_1(w-1)}.$$

The case $(q,w) = (3,3)$ has been investigated by Svanström [11], who established the following.

*Lemma 13 (Svanström [11]):*

$$A_3(n,4,[2,1]) = \begin{cases} \frac{n(n-2)}{4}, & \text{if } n \text{ is even} \\ \frac{n(n-1)}{4} - \frac{\epsilon(n)}{6}, & \text{if } n \in \{7,11\}. \end{cases}$$

The value of $A_3(n,4,[2,1])$ remains undetermined for all odd $n \notin \{7,11\}$. We address this problem in what follows.

TABLE II
WILSON'S FUNDAMENTAL CONSTRUCTION

| | |
|---|---|
| Input: | (master) GDD $\mathcal{D} = (X, \mathcal{G}, \mathcal{A})$; |
| | weight function $\omega : X \to \mathbb{Z}_{\geq 0}$; |
| | (ingredient) $K$-GDD $\mathcal{D}_A = (X_A, \mathcal{G}_A, \mathcal{B}_A)$ of type $[\omega(a) : a \in A]$ for each block $A \in \mathcal{A}$, where |
| | $X_A = \cup_{a \in A}\{\{a\} \times \{1,\ldots,\omega(a)\}\}$, and |
| | $\mathcal{G}_A = \{\{a\} \times \{1,\ldots,\omega(a)\} : a \in A\}$. |
| Output: | $K$-GDD $\mathcal{D}^* = (X^*, \mathcal{G}^*, \mathcal{A}^*)$ of type $[\sum_{x \in G} \omega(x) : G \in \mathcal{G}]$, where |
| | $X^* = \cup_{x \in X}(\{x\} \times \{1,\ldots,\omega(x)\})$, |
| | $\mathcal{G}^* = \{\cup_{x \in G}(\{x\} \times \{1,\ldots,\omega(x)\}) : G \in \mathcal{G}\}$, and |
| | $\mathcal{A}^* = \cup_{A \in \mathcal{A}} \mathcal{B}_A$. |
| Notation: | $\mathcal{D}^* = \text{WFC}(\mathcal{D}, \omega, \{\mathcal{D}_A : A \in \mathcal{A}\})$. |
| Note: | By convention, for $x \in X$, $\{x\} \times \{1,\ldots,\omega(x)\} = \varnothing$ if $\omega(x) = 0$. |

*A. Optimal $(n,4,[2,1])_3$-Codes: The Case $n \equiv 1 \pmod 4$*

We know from Lemma 7 that $A_3(n,4,[2,1]) = n(n-1)/4$ for $n \in \{5,9,13\}$. Hence, using Theorem 3, we obtain $A_3(n,4,[2,1]) = n(n-1)/4$ for all $n \in B(\{5,9,13\})$. The PBD-closure of $\{5,9,13\}$ has been determined by Hamel et al. [27].

*Theorem 7 (Hamel et al. [27]):* $B(\{5,9,13\}) = \{n : n \equiv 1 \pmod 4\} \setminus \{17,29,33\}$.

*Corollary 1:* $A_3(n,4,[2,1]) = n(n-1)/4$ for all $n \equiv 1 \pmod 4$.

*Proof:* The case $n \notin \{17,29,33\}$ follows from Theorem 7. Optimal $(n,4,[2,1])_3$-codes for $n \in \{17,29,33\}$ are provided by Lemma 7. □

This solves the problem of optimal $(n,4,[2,1])_3$-codes completely for the case of $n \equiv 1 \pmod 4$.

*B. Optimal $(n,4,[2,1])_3$-Codes: The Case $n \equiv 3 \pmod 4$*

This case seems considerably more difficult. More tools and additional results are required. We first consider the case $n \equiv 3 \pmod{12}$.

*Lemma 14:* There exists a PBD$(n,\{21,33,41\})$ for all sufficiently large $n \equiv 1 \pmod 4$.

*Proof:* $\alpha(\{21,33,41\}) = 4$ and $\beta(\{21,33,41\}) = 4$. So by Theorem 4, there exists a PBD$(n,\{21,33,41\})$ for all sufficiently large $n$ satisfying $n - 1 \equiv 0 \pmod 4$ and $n(n-1) \equiv 0 \pmod 4$. These congruences are satisfied if and only if $n \equiv 1 \pmod 4$. □

*Lemma 15:* There exists a $\{5,9\}$-GDD of type $3^u$ for all sufficiently large $u \equiv 1 \pmod 4$.

*Proof:* The set

$$U(K,g) = \{u : \exists \text{ a } K\text{-GDD of type } g^u\}$$

where $K$ and $g$ are fixed, is well known to be PBD-closed (see, for example, [28]). We know $21, 33, 41 \in U(\{5,9\},3)$ from Theorem 6 and Lemma 2. So $B(\{21,33,41\}) \subseteq U(\{5,9\},3)$. The required result now follows from Lemma 14. □

*Corollary 2:* $A_3(n,4,[2,1]) = n(n-1)/4 - n/6$ for all sufficiently large $n \equiv 3 \pmod{12}$.

*Proof:* Take a $\{5,9\}$-GDD of type $3^{n/3}$, which exists by Lemma 15, and apply Theorem 2 with $d = 4$ and $\bar{w} = [2,1]$. The $(n,4,[2,1])_3$-code $\mathcal{C}$ constructed has size

$$|\mathcal{C}| \geq 5b_5 + 18b_9 + n/3. \quad (4)$$

Next note that since every pair of points not contained in a group appears in exactly one block of a GDD, we have

$$b_5 \binom{5}{2} + b_9 \binom{9}{2} = \binom{n}{2} - n. \quad (5)$$

From (4) and (5), we derive

$$|\mathcal{C}| \geq \frac{n(n-1)}{4} - \frac{n}{6}.$$

The matching upper bound is provided by Lemma 3. □

We now consider the case $n \equiv 7$ or $11 \pmod{12}$. We begin with some definitions. A *transversal design* TD$(k,g)$ is a $\{k\}$-GDD of type $g^k$, and is equivalent to the existence of $k-2$ mutually orthogonal Latin squares of side $g$.

*Theorem 8 (Chowla et al. [29]):* For any fixed $k$, there exists a constant $g_0(k)$ such that a TD$(k,g)$ exists for all $g \geq g_0(k)$.

We use Wilson's Fundamental Construction [20], described in Table II, to construct GDDs that we require.

*Theorem 9:* There exist $\{5,9\}$-GDDs of types $3^u 7^1$ and $3^u 11^1$ for all sufficiently large $u \equiv 0 \pmod 4$.

*Proof:* Let $g \equiv 0 \pmod 3$ be sufficiently large so that a TD$(11,g)$ exists by Theorem 8. Take this TD$(11,g)$ as the master GDD in Wilson's Fundamental Construction and assign weight four to all points in nine groups and assign weight zero or four to the points in the other two groups. The ingredient GDDs are $\{5,9\}$-GDDs of types $4^9, 4^{10}$ and $4^{11}$. These ingredient GDDs all exist.

1) There exists a $\{5\}$-GDD of type $9^1 1^{28}$ [27]. Delete from this GDD a point $p$ not in the group of size nine. By taking the blocks through $p$ (with $p$ deleted) as groups and the group of size nine as a block gives the required $\{5,9\}$-GDD of type $4^9$.
2) The $\{5,9\}$-GDDs of types $4^{10}$ and $4^{11}$ are provided by Theorem 6.

Hence, Wilson's Fundamental Construction gives a $\{5,9\}$-GDD of type $(4g)^9(4x)^1(4y)^1$ for $0 \leq x, y \leq g$.

Let $x \equiv 0 \pmod 3$ and $y \in \{1, 2\}$. Now adjoin three ideal points to this GDD and fill in the groups with $\{5,9\}$-GDDs of types $3^{(4g+3)/3}$ and $3^{(4x+3)/3}$. These GDDs all exist by Lemma 15 since we can always choose $g$ and $x$ to be sufficiently large. By aligning the three ideal points with the last group, we obtain a $\{5,9\}$-GDD of type $3^{12g+4x/3}(4y+3)^1$. Furthermore, we can always choose $4x/3 \in \{g, g+3, g+6, g+9\}$ if $g \geq 27$. The theorem now follows. □

*Corollary 3:* $A_3(n, 4, [2, 1]) = n(n-1)/4 - \epsilon(n)/6$ for all sufficiently large $n \equiv 7$ or $11 \pmod{12}$.

*Proof:* The upper bound on $A_3(n, 4, [2, 1])$ is given by Lemma 3. For the lower bound, take $\{5, 9\}$-GDDs of types $3^{(n-7)/3}7^1$ and $3^{(n-11)/3}11^1$, which exist by Theorem 9, and apply Theorem 2 with $d = 4$ and $\bar{w} = [2, 1]$. It can be checked in a manner similar to the proof of Corollary 2 that the $(n, 4, [2, 1])_3$-codes constructed have the required size. □

We summarize the results of this section as follows.

*Theorem 10:* $A_3(n, 4, [2, 1]) = n(n-1)/4 - \epsilon(n)/6$ for all odd $n$ sufficiently large.

As a consequence, the exact value of $A_3(n, 4, [2, 1])$ is now known for all $n$ sufficiently large.

### C. Some Optimal $(n, 6, \bar{w})_q$-Codes

Svanström *et al.* [12] showed that $A(10, 6, [3, 1]) = 10$ and $A(10, 6, [2, 2]) = 15$. Hence, using Theorem 3, we obtain $A_3(n, 6, [3, 1]) = n(n-1)/9$ and $A_3(n, 6, [2, 2]) = n(n-1)/6$ for all $n \in B(\{10\})$.

*Lemma 16:* For all sufficiently large $n \equiv 1$ or $10 \pmod{45}$

$$A_3(n, 6, [3, 1]) = n(n-1)/9$$
$$A_3(n, 6, [2, 2]) = n(n-1)/6$$
$$A_4(n, 6, [2, 1, 1]) = n(n-1)/6.$$

*Proof:* By Theorem 4, $B(\{10\})$ contains all sufficiently large integers $n \equiv 1$ or $10 \pmod{45}$. So the values of $A_3(n, 6, [3, 1])$ and $A_3(n, 6, [2, 2])$ are as claimed. For the value of $A_4(n, 6, [2, 1, 1])$, note that $[2, 1, 1]$ is a refinement of $[2, 2]$ and apply Lemma 6. So $A_4(n, 6, [2, 1, 1]) \geq n(n-1)/6$. The corresponding upper bound is from Lemma 1. □

### VII. OPTIMAL CONSTANT-COMPOSITION CODES OF WEIGHT $w$ AND DISTANCE $2w - 3$

Let $\bar{w} = [w_1, \ldots, w_{q-1}]$, $w = \sum_{i=1}^{q-1} w_i$, and $d = 2w - 3$. Lemma 1 gives

$$A_q(n, d, \bar{w}) \leq \left\lfloor \frac{n}{w_1} \left\lfloor \frac{n-1}{w_1 - 1} \right\rfloor \right\rfloor.$$

So when $n$ satisfies

$$n - 1 \equiv 0 \pmod{w_1 - 1}$$

and

$$n(n-1) \equiv 0 \pmod{w_1(w_1 - 1)}$$

we have

$$A_q(n, d, \bar{w}) \leq \frac{n(n-1)}{w_1(w_1 - 1)}.$$

*Lemma 17:* $A_3(n, 5, [3, 1]) = n(n-1)/6$ for all sufficiently large $n \equiv 1$ or $3 \pmod{6}$.

*Proof:* Lemma 4 gives $A_3(n, 5, [3, 1]) \leq n(n-1)/6$. So we only have to establish the lower bound. Lemma 10 gives $A_3(n, 5, [3, 1]) = n(n-1)/6$ for $n \in \{7, 13, 15, 19\}$. By Theorem 3, $A_3(n, 5, [3, 1]) = n(n-1)/6$ for all $n \in B(\{7, 13, 15, 19\})$. Now, $\alpha(\{7, 13, 15, 19\}) = 2$ and $\beta(\{7, 13, 15, 19\}) = 6$. It follows from Theorem 4 that $B(\{7, 13, 15, 19\})$ contains all sufficiently large $n$ satisfying $n - 1 \equiv 0 \pmod 2$ and $n(n-1) \equiv 0 \pmod 6$, which are satisfied if and only if $n \equiv 1$ or $3 \pmod 6$. □

TABLE III
ELEMENTS THAT ARE NOT IN OR NOT KNOWN TO BE IN $B(\{4, 7, 8, 9\})$

| 5 | 6 | 10 | 11 | 12 | 14 | 15 | 17 | 18 | 19 |
|---|---|---|---|---|---|---|---|---|---|
| 20 | 21 | 23 | 24 | 26 | 27 | 30 | 35 | 38 | 39 |
| 41 | 42 | 44 | 47 | 48 | 51 | 54 | 59 | 62 | 110 |
| 143 | 146 | 147 | 150 | 158 | 159 | 161 | 162 | 164 | 167 |
| 170 | 171 | 173 | 174 | | | | | | |

TABLE IV
ELEMENTS THAT ARE NOT IN OR NOT KNOWN TO BE IN $B(\{8, 9, 10\})$ ($a$-$b$ DENOTES THE NUMBERS $a, a+1, \ldots, b$)

| 11–56 | 58–63 | 66–71 | 75–79 | 101–109 | 111–113 |
|---|---|---|---|---|---|
| 115–119 | 126–127 | 133–135 | 155–160 | 166–167 | 173–231 |
| 239 | 247–287 | 290–295 | 299–343 | 346–351 | 355–399 |
| 403–407 | 411–423 | 426–431 | 435–439 | 443–448 | 452–455 |
| 472–497 | 499–503 | 507–511 | 580–582 | | |

*Lemma 18:* $A_3(n, 7, [4, 1]) = n(n-1)/12$ for all sufficiently large $n \equiv 1$ or $4 \pmod{12}$.

*Proof:* Lemma 4 gives $A_3(n, 7, [4, 1]) \leq n(n-1)/12$. So we only have to establish the lower bound. Lemma 11 gives $A_3(n, 7, [4, 1]) = n(n-1)/12$ for $n = 13$ and $n = 16$. By Theorem 3, $A_3(n, 7, [4, 1]) = n(n-1)/12$ for all $n \in B(\{13, 16\})$. Now, $\alpha(\{13, 16\}) = 3$ and $\beta(\{13, 16\}) = 12$. It follows from Theorem 4 that $B(\{13, 16\})$ contains all sufficiently large $n$ satisfying $n - 1 \equiv 0 \pmod 3$ and $n(n-1) \equiv 0 \pmod{12}$, which are satisfied if and only if $n \equiv 1$ or $4 \pmod{12}$. □

### VIII. OPTIMAL QUATERNARY CONSTANT-COMPOSITION CODES OF WEIGHT THREE

#### A. The Case of Distance Three

From Lemma 5, $A_4(n, 3, [1, 1, 1]) \leq n(n-1)$ for all $n$. We also know from Lemma 8 that $A_4(n, 3, [1, 1, 1]) = n(n-1)$ for $n \in \{4, 7, 8, 9, 10\}$. Hence, using Theorem 3, we obtain $A_4(n, 3, [1, 1, 1]) = n(n-1)$ for all $n \in B(\{4, 7, 8, 9\}) \cup B(\{8, 9, 10\})$. The PBD-closures of $\{4, 7, 8, 9\}$ and $\{8, 9, 10\}$ have been determined by Mullin *et al.* [30], and Colbourn and Ling [31], respectively.

*Theorem 11 (Mullin et al. [30]):* A PBD$(n, \{4, 7, 8, 9\})$ exists for all integers $n \geq 4$ and $n$ not in Table III.

*Theorem 12 (Colbourn and Ling [31]):* A PBD$(n, \{8, 9, 10\})$ exists for all integers $n \geq 8$ and $n$ not in Table IV.

*Lemma 19:* The following GDDs exist:
  i) $\{4\}$-GDD of type $9^4 12^1$;
  ii) $\{4\}$-GDD of type $36^4 15^1$;
  iii) $\{4\}$-GDD of type $36^4 30^1$.
  *Proof:* Follows from Theorem 5. □

Combining the above results with Lemma 12 gives the following.

*Corollary 4:* $A_4(n, 3, [1, 1, 1]) = n(n-1)$, for all $n \geq 4$, except for $n \in \{5, 6\}$ when we have $A_4(5, 3, [1, 1, 1]) = 18$

and $A_4(6, 3, [1, 1, 1]) = 28$, and except possibly for $n \in \{44, 47, 51, 54, 59, 62, 158, 167, 173\}$.

*Proof:* By Theorems 11 and 12, $A_4(n, 3, [1, 1, 1]) = n(n-1)$, for all $n \geq 4$, except for

$$n \in \{5, 6, 11, 12, 14, 15, 17, 18, 19, 20, 21, 23, 24, 26,$$
$$27, 30, 35, 38, 39, 41, 42, 44, 47, 48, 51,$$
$$54, 59, 62, 158, 159, 167, 173, 174\}.$$

The value of $A_4(n, 3, [1, 1, 1])$ for $n = 5$ and $n = 6$ is determined by Lemma 12, while Lemma 8 determines the value of $A_4(n, 3, [1, 1, 1])$ for

$$n \in \{11, 12, 14, 15, 17, 18, 19, 20, 21, 23,$$
$$24, 26, 27, 30, 35, 38, 39, 41, 42\}.$$

The value of $A_4(n, 3, [1, 1, 1])$ for $n \in \{48, 159, 174\}$ follows from Lemma 19 and Theorem 2. □

### B. The Case of Distance Four

We know from Lemma 5 that $A_4(n, 4, [1, 1, 1]) \leq n(n-1)/2$ when $n$ is odd. We also know from Lemma 9 that $A_4(n, 4, [1, 1, 1]) = n(n-1)/2$ when $n \in \{11, 19\}$. Hence, using Theorem 3, we obtain $A_4(n, 4, [1, 1, 1]) = n(n-1)/2$ for all $n \in B(\{11, 19\})$. The PBD-closure of $\{11, 19\}$ has not been determined precisely, but Theorem 4 gives the following.

*Lemma 20:* A $PBD(n, \{11, 19\})$ exists for all sufficiently large integers $n \equiv 1 \pmod 2$.

The following is now immediate.

*Corollary 5:* For all sufficiently large integers $n \equiv 1 \pmod 2$, $A_4(n, 4, [1, 1, 1]) = n(n-1)/2$.

*Lemma 21:* For all sufficiently large integers $n \equiv 0 \pmod 2$, $A_4(n, 4, [1, 1, 1]) = n(n-2)/2$.

*Proof:* For $n$ even, $A_4(n, 4, [1, 1, 1]) \leq n(n-2)/2$ is given by Lemma 5, so we just need to exhibit a matching lower bound for $n$ large enough.

Let $n$ be large enough so that an optimal $(n+1, 4, [1, 1, 1])_4$-code exists. This code has size $(n+1)n/2$ by Corollary 5. For each position $r$, any two codewords agreeing in a nonzero value in position $r$ must have their remaining nonzero positions all different from each other. Hence, there are at most $n/2$ of these. Since a nonzero value in position $r$ can take on three different values, the total number of codewords having nonzero value in position $r$ is at most $3n/2$. Therefore, shortening an optimal $(n+1, 4, [1, 1, 1])_4$-code in any position results in an $(n, 4, [1, 1, 1])_4$-code of size at least

$$\frac{(n+1)n}{2} - \frac{3n}{2} = \frac{n(n-2)}{2}. \qquad \square$$

### C. The Case of Distance Five

*Lemma 22:*
$$A_4(n, 5, [1, 1, 1]) = \begin{cases} 1, & \text{if } n \leq 4 \\ 2, & \text{if } n = 5 \\ 4, & \text{if } n = 6 \\ n, & \text{if } n \geq 7. \end{cases}$$

*Proof:* The cases $n \leq 6$ can be easily verified. For $n \geq 7$, taking

$$12030^{n-4}$$

together with all its cyclic shifts gives the $n$ codewords of an $(n, 5, [1, 1, 1])_4$-code. This together with Lemma 5 proves the required result. □

## IX. SUMMARY

The following summarizes the key results obtained in this paper.

*Theorem 13:* For all sufficiently large $n$,
 i)
$$A_3(n, 4, [2, 1])$$
$$= \begin{cases} \left\lfloor \frac{n}{2} \left\lfloor \frac{n-1}{2} \right\rfloor \right\rfloor, & \text{if } n \equiv 0, 1, \text{ or } 2 \pmod 4 \\ \frac{(n-1)^2}{4} + \left\lfloor \frac{n-3}{12} \right\rfloor, & \text{if } n \equiv 3 \pmod 4. \end{cases}$$

 ii)
$$A_4(n, 3, [1, 1, 1]) = n(n-1).$$

 iii)
$$A_4(n, 4, [1, 1, 1]) = n \left\lfloor \frac{n-1}{2} \right\rfloor.$$

 iv)
$$A_4(n, 5, [1, 1, 1]) = n.$$

It follows that the size of constant-composition codes of weight three is now determined for all sufficiently large lengths. Previously, only $A_3(n, 4, [2, 1])$ was determined for $n$ even [11].

## X. CONCLUSION

In this paper, we established an interesting PBD-closure result for the set of lengths $n$ of constant-composition codes having size of the form $cn(n-1)$, for some constant $c$, provided its distance is not too small. As a consequence of Wilson's theory of PBD-closure, the size of optimal constant-composition codes can be determined for infinitely many $n$ from just an example of an optimal code. More precise constructions based on group divisible designs are also given, which enabled us to determine the size of several families of optimal constant-composition codes. In particular, the size of optimal constant-composition codes of weight three is determined for all lengths sufficiently large.

The main purpose of this paper was to introduce the approach of PBD-closure as a technique for determining the size of optimal constant-composition codes. The reader may notice the following:
  i) the size of optimal codes are determined only for sufficiently large lengths, and
  ii) coverage of the application of our approach is not comprehensive.

First, it is indeed possible to derive concrete bounds on the length of codes, and often even we are able to derive a specific finite set of possible exceptions, for which the size of optimal codes can be determined. However, such derivations are usually highly technical, require deep methods in combinatorial design theory, and would present a distraction from the main method of this paper if pursued. Second, it is impossible to be comprehensive in covering the application of our approach due to its general nature. We have illustrated this by determining the size of some families of optimal constant-composition codes. Moreover, the technique does not only work for optimal constant-composition codes. It can also be applied to the construction of "good" nonoptimal constant-composition codes. We are certain that the approach taken in this paper will yield exact determination or better lower bounds of $A_q(n, d, \bar{w})$ for more parameter sets through the construction of larger classes of group divisible designs and the discovery of more optimal constant-composition codes.

ACKNOWLEDGMENT

Part of this work was performed while H. Shen was visiting the Division of Mathematical Sciences, School of Physical and Mathematical Sciences, Nanyang Technological University, Singapore. He is grateful to the institution for its hospitality.